# Combined photon–proton modeling of radiation-induced brain imaging changes confirms variability in proton relative biological effectiveness and increased periventricular radiosensitivity

Martina Palkowitsch[1,2], Larissa S. Kilian[1], Fabian Hennings[1,2], Armin Lühr[3,4], Justus Thiem[1,3,5], Arne Grey[6], Rebecca Bütof[1,2,3,5], Annekatrin Seidlitz[1,3,5], Esther G.C. Troost[1,2,3,5,7], Mechthild Krause[1,2,3,5,7], Steffen Löck[1,2,3,5,7]

[1]OncoRay – National Center for Radiation Research in Oncology, Faculty of Medicine and University Hospital Carl Gustav Carus, TUD Dresden University of Technology, Helmholtz-Zentrum Dresden-Rossendorf, Dresden, Germany.
[2]Helmholtz-Zentrum Dresden-Rossendorf, Institute of Radiooncology – OncoRay, Dresden, Germany.
[3]Department of Radiotherapy and Radiation Oncology, Faculty of Medicine and University Hospital Carl Gustav Carus, TUD Dresden University of Technology, Dresden, Germany.
[4]TU Dortmund University, Department of Physics, Dortmund, Germany.
[5]National Center for Tumor Diseases (NCT), NCT/UCC Dresden, a partnership between DKFZ, Faculty of Medicine and University Hospital Carl Gustav Carus, TUD Dresden University of Technology, and Helmholtz-Zentrum Dresden-Rossendorf (HZDR), Germany.
[6]Institute of Neuroradiology, Faculty of Medicine and University Hospital Carl Gustav Carus, Technische Universität Dresden, Dresden, Germany.
[7]German Cancer Consortium (DKTK), Partner Site Dresden, and German Cancer Research Center (DKFZ), Heidelberg, Germany.

**Corresponding author:**
Martina Palkowitsch
E-mail: martina.palkowitsch@oncoray.de
OncoRay – National Center for Radiation Research in Oncology
Fetscherstr. 74, 01307 Dresden, Germany

*Abbreviations*: AUC, area under the receiver operating characteristic curve; $AUC_{mean}$, mean AUC; CTV, clinical target volume; D, absorbed dose; $D_{constRBE}$, D weighted by a constant RBE (1.0 for photons and 1.1 for protons); DS, double scattering; fuMRI, follow-up MRI; DVH, dose-volume histogram; $D_x$, minimum dose received by the highest-dose x% of the volume; $D_{xml}$, minimum dose received by the highest-dose x ml of the volume; IMRT, intensity-modulated radiotherapy; LET, linear energy transfer; $LET_d$, dose-averaged LET; MC, Monte Carlo; MRI, magnetic resonance imaging; NTCP, normal tissue complication probability; PBS, pencil beam scanning; pCT, planning computed tomography; PVR, periventricular region; RBE, relative biological effectiveness; RICE, radiation-induced contrast enhancement; $t_{RICE}$, time from the start of radio(chemo)therapy to first RICE detection; $V_x$ volume receiving more than x Gy(RBE); 3D-CRT, three-dimensional conformal radiotherapy.

## Abstract

**Purpose:** Recent proton-only investigations of radiation-induced contrast enhancements (RICE) after brain tumor radiotherapy indicated clinically relevant variability in proton relative biological effectiveness (RBE) and increased radiosensitivity of the periventricular region (PVR). Because RBE is defined relative to a reference radiation, these studies required assumptions on the photon dose-response relationship. Integrating photon-treated patients provides an explicit clinical reference, enabling separation of modality-independent dose and PVR effects from proton-specific effects related to linear energy transfer (LET). This study aimed to validate LET-dependent proton RBE variability and PVR radiosensitivity using cross-modality, spatially resolved predictive modeling of RICE in a combined photon-proton cohort.

**Methods and Materials:** Predictive models for RICE detected on follow-up magnetic resonance imaging were developed in 152 intracranial tumor patients treated with photons (n = 65) or protons (n = 87). Logistic regression was applied at the voxel level to model spatial occurrence and at the patient level to model incidence. Performance was quantified by the mean area under the receiver operating characteristic curve ($AUC_{mean}$) from cross-validation. A clinical RBE model was derived from voxel-wise comparisons of estimated risk between photon and proton irradiation.

**Results:** In total, 128 RICE of various grades occurred in 64 patients. Voxel-level modeling identified absorbed dose (D), D multiplied by dose-averaged LET ($LET_d$) for proton therapy, and PVR as independent predictors of RICE ($AUC_{mean}$ = 0.89 [95% confidence interval: 0.82-0.97]). The model implied a variable proton RBE described by $RBE = 1 + m \cdot LET_d$, with $m = 0.10$ µm/keV. At the patient level, PVR $D_{2ml}$ based on this RBE achieved the highest predictive performance (0.78 [0.67-0.89]).

**Conclusions:** RICE was spatially associated with dose and PVR proximity across photon and proton radiotherapy, with an additional LET-dependent component after proton therapy. The cross-modality framework validates proton RBE variability against an observed photon reference rather than predefined reference assumptions, and supports PVR radiosensitivity as modality-independent. Accounting for variable proton RBE and the PVR as an organ at risk may improve risk assessment and mitigation of radiation-induced side effects.

## Introduction

Protons interact with matter differently than photons, resulting in two key characteristics: highly localized dose deposition at the Bragg peak and differences in biological effectiveness for a given absorbed dose. To translate dose prescriptions from photon to proton therapy, photon doses are scaled by the relative biological effectiveness (RBE) to obtain isoeffective proton doses [1]. Clinically, a constant RBE of 1.1 is widely adopted [2–4], although RBE is known to vary with multiple physical and biological parameters, including linear energy transfer (LET) [5].
As protons decelerate, LET increases, leading to denser ionization patterns, more complex DNA damage, and elevated RBE toward the distal edge of the Bragg peak [5,6]. This results in an effective extension of the biological range beyond the physical range [7–9]. Together with safety margins for range and setup uncertainties, this raises concern that biological effectiveness in adjacent normal tissue may be underestimated [2,3,10,11]. Consequently, the adequacy of a constant proton RBE has been questioned, particularly with regard to risk of unanticipated toxicities. In response, LET-informed optimization strategies that explicitly account for RBE variability have been proposed and shown to reduce estimated toxicity risk in treatment planning studies [11–19]. However, their routine clinical implementation requires a more comprehensive understanding of proton RBE variability in patients [20].
Despite strong experimental evidence for LET-dependent RBE, its clinical relevance for normal tissue toxicity remains difficult to establish [21,22]. RBE effects are spatially localized and largely confined to the distal millimeters of the proton range. Moreover, heterogeneous LET and dose distributions, particularly within organs at risk, complicate the interpretation of clinical toxicity outcomes. Consequently, conventional population-based normal tissue complication probability (NTCP) models, which rely on organ-level dose metrices, may lack sensitivity to detect such localized effects [21,23].
Spatially resolved analyses provide a potential solution by enabling voxel-wise correlation of radiation response with underlying dose and LET distributions [21]. An increasingly investigated endpoint in this context is radiation-induced contrast enhancement (RICE) observed on follow-up magnetic resonance imaging (MRI) after brain tumor radiotherapy [21,24–39]. Although often asymptomatic, RICE can progress to clinically significant symptoms [40,41]; in a recent study of glioma patients, 26% developed severe symptoms requiring treatment [41]. Furthermore, RICE remains difficult to distinguish from tumor recurrence, thereby complicating patient management [24].
Although similar imaging changes are observed after photon therapy, recent evidence suggests that uncertainties in proton biological effectiveness may represent an important risk factor [41]. Proton-only studies have demonstrated that RICE risk can be predicted using absorbed physical dose (D), D multiplied by dose-averaged LET ($LET_d$), and proximity to the cerebral ventricles [32,34,36,37,42]. These models enabled the derivation of LET-dependent clinical RBE models; but, because RBE is defined relative to a reference radiation, absence of photon data required assumptions about the reference dose-response relationship.
A combined photon–proton analysis overcomes this limitation by enabling direct comparison of dose–response relationships across radiation modalities. It allows shared dose and anatomical effects to be separated from proton-specific LET effects by using photon-treated patients as a clinical reference, while testing whether the radiosensitivity of the periventricular region (PVR) is also observed after photon therapy.
Accordingly, this study aims to (i) develop predictive models of RICE in a combined photon–proton cohort and derive a variable RBE model using photon data as reference; and (ii) evaluate whether the PVR constitutes a risk factor for RICE also in the photon cohort.

## Materials and methods

**Patient cohorts**

In this retrospective study, we analyzed a cohort of 152 adult patients (≥ 18 years at time of treatment) with primary intracranial tumors treated with either photon (n = 65) or proton therapy (n = 87) at the University Proton Therapy Dresden and the Department of Radiotherapy and Radiation Oncology, University Hospital Carl Gustav Carus Dresden, between 2015 and 2019. Patients were consecutively included. Subsets of this cohort were included in previous publications [37,41,42]. Exclusion criteria were prior cranial irradiation, treatment with combined proton–photon therapy, and missing or inadequate follow-up MRI or incomplete dose and LET simulation datasets. The cohort included patients with intracranial tumors of various histologies and demographic characteristics. Photon therapy was delivered using three-dimensional conformal radiotherapy (3D-CRT; n = 32) or intensity-modulated radiotherapy (IMRT; n = 33). Proton therapy was delivered using double scattering (DS) for treatments initiated before 2018 (n = 61) and pencil beam scanning (PBS) for treatments initiated in 2018 or later (n = 26). Most patients received a prescribed dose of either 54 or 60 Gy(RBE) in 2 Gy(RBE) fractions, delivered to one or two target volumes. An RBE of 1.0 was assumed for photon and 1.1 for proton therapy delivery. The study was approved by the Ethics Committee of the Dresden University of Technology, Germany (SR+BO-EK-252062022). Patient, tumor, and treatment characteristics for the different cohorts are summarized in Table 1 and Supplementary Tables S1-S4.

**Dose and LET calculation**

For proton patients treated with DS, clinical treatment planning was performed using the CMS XiO treatment planning system (TPS; Elekta, Stockholm, Sweden). For PBS patients, the RayStation TPS (RaySearch Laboratories AB, Stockholm, Sweden) was used. For all proton patients, Monte Carlo (MC)–based dose and $LET_d$ distributions were retrospectively calculated. DS patients were simulated using an in-house MC framework based on TOPAS [43–45], and PBS patients using RayStation (research version v8.99.30.101). In all cases, $LET_d$ was defined as unrestricted proton $LET_d$ in unit-density tissue, in accordance with current recommendations [10,46]. For patients with multiple treatment plans (e.g., a boost), the voxel-wise total dose was calculated as the sum of doses from all plans. The corresponding $LET_d$ was calculated as the dose-weighted average across plans, obtained by summing the product of dose and $LET_d$ for each plan and dividing by the total dose. For photon therapy, the clinically planned dose distribution was analyzed, and a constant LET of 0.31 keV/μm was assumed to represent secondary electron interactions in accordance with the literature [47–50].

**RICE diagnosis, image registration and segmentation**

All patients underwent follow-up MRI approximately every three months to assess tumor recurrence and RICE. RICE was defined as new contrast-enhancing lesions not attributable to tumor recurrence. Diagnosis and classification were based on contrast-enhanced T1-weighted follow-up MRIs (fuMRIs). A board-certified neuroradiologist identified RICE as areas of new or progressive contrast enhancement. Ambiguous cases were additionally evaluated in consultation with a senior, board-certified neuroradiologist. Classification relied either on histological confirmation of post-radiogenic tissue injury or on spontaneous partial or complete regression of the enhancement without changes in therapy. After RICE diagnosis, lesions were retrospectively identified on the earliest fuMRI at which they were visible.

**Table 1. Patient characteristics.** Two-sided p-values were calculated to assess differences between patients with and without radiation-induced contrast enhancement (RICE). The $\chi^2$ test was used for categorical variables and the Mann–Whitney-U test for continuous variables.

| Variables | All patients | | Non-RICE patients | | RICE patients | | |
|---|---|---|---|---|---|---|---|
| | n | (%) | n | (%) | n | (%) | p-value |
| Number of patients | 152 | (100) | 88 | (58) | 64 | (42) | |
| Treatment modality | | | | | | | 0.034 |
| Photon | 65 | (43) | 44 | (68) | 21 | (32) | |
| Proton | 87 | (57) | 44 | (51) | 43 | (49) | |
| Treatment period | 2012-2019 | | 2012-2019 | | 2012-2019 | | |
| | Median | (range) | Median | (range) | Median | (range) | p-value |
| Age | 50 | (24-82) | 48 | (24-79) | 52 | (27-82) | 0.25 |
| Clinical target volume (cm³) | 191 | (5-569) | 169 | (5-501) | 219 | (26-569) | < 0.001 |
| Prescribed dose (Gy (RBE)) | 60 | (54-60) | 60 | (54-60) | 60 | (54-60) | < 0.001 |
| Dose per fraction (Gy(RBE)) | 2.0 | (1.8-2.0) | 2.0 | (1.8-2.0) | 2.0 | (2.0-2.0) | 0.14 |
| | n | (%) | n | (%) | n | (%) | p-value |
| Treatment technique | | | | | | | 0.17 |
| 3D-CRT | 32 | (21) | 23 | (72) | 9 | (28) | |
| IMRT | 33 | (22) | 21 | (64) | 12 | (36) | |
| Double scattering | 61 | (40) | 30 | (49) | 31 | (51) | |
| Pencil beam scanning | 26 | (17) | 14 | (54) | 12 | (46) | |
| Sex | | | | | | | 0.12 |
| Female | 80 | (53) | 51 | (64) | 29 | (36) | |
| Male | 72 | (47) | 37 | (51) | 35 | (49) | |
| Histology | | | | | | | < 0.001 |
| Astrocytoma | 34 | (22) | 21 | (62) | 13 | (38) | |
| Craniopharyngioma | 1 | (1) | 1 | (100) | 0 | (0) | |
| Ependymoma | 4 | (3) | 4 | (100) | 0 | (0) | |
| Glioblastoma | 51 | (34) | 18 | (35) | 33 | (65) | |
| Glioma | 2 | (1) | 2 | (100) | 0 | (0) | |
| Hemangiopericytoma | 1 | (1) | 0 | (0) | 1 | (100) | |
| Meningioma | 23 | (15) | 21 | (91) | 2 | (9) | |
| Oligoastrocytoma | 16 | (10) | 13 | (81) | 3 | (19) | |
| Oligodendroglioma | 17 | (11) | 5 | (29) | 12 | (71) | |
| Pituitary adenoma | 1 | (1) | 1 | (100) | 0 | (0) | |
| Unkown | 2 | (1) | 2 | (100) | 0 | (0) | |
| WHO tumor grade | | | | | | | < 0.001 |
| 1 | 18 | (12) | 17 | (94) | 1 | (6) | |
| 2 | 21 | (14) | 12 | (57) | 9 | (43) | |
| 3 | 58 | (38) | 38 | (66) | 20 | (34) | |
| 4 | 52 | (34) | 19 | (37) | 33 | (63) | |
| n.g. | 3 | (2) | 2 | (67) | 1 | (33) | |
| Tumor resection | | | | | | | 0.056 |
| Yes | 122 | (80) | 66 | (54) | 56 | (46) | |
| No | 30 | (20) | 22 | (73) | 8 | (27) | |

*Abbreviations:* 3D-CRT, three-dimensional conformal radiation therapy; DS, double scattering; IMRT, intensity-modulated radiotherapy; PBS, pencil beam scanning; RBE, relative biological effectiveness; RICE, radiation-induced contrast enhancement.

Since RICE tend to grow, this approach reduced uncertainty in lesion onset localization. RICE were manually contoured in RayStation (v14.10.100.0). The resulting contours were transferred from the fuMRI to the planning computed tomography (pCT) using non-linear image registration.

All image data were processed using a customized analysis pipeline as described by Eulitz et al. [37]. In brief, anatomical MRIs were first bias-corrected and resliced to an isotropic voxel size of 1 mm. Tissue segmentation was then performed on the treatment-planning MRI using an atlas-based method to delineate the ventricular system, cerebrospinal fluid, and white matter. The PVR was defined as a 4 mm margin surrounding the ventricles. The resulting segmentations were subsequently non-linearly registered to the pCT. All further analyses were performed in pCT space.

**Risk analysis and modelling**

Patient characteristics between patients with and without RICE were compared using the Mann-Whitney U test for continuous variables and the $\chi^2$ test for categorical variables. A two-sided p-value < 0.05 was considered statistically significant.

Risk factors associated with the spatial location of RICE and with RICE occurrence were analyzed using voxel-level and patient-level models following the methodology described in previous studies [36,37]. Logistic regression analyses were performed in the following cohorts: (i) combined proton–photon cohort, (ii) photon cohort (ii) proton cohort (DS + PBS), (iii) proton DS cohort, and (iv) proton PBS cohort.

For each cohort, voxel-level and patient-level models were constructed using different sets of input variables. Model performance was evaluated using repeated 3-fold cross-validation with 333 repetitions. For each cohort, the best-performing model was selected based on the mean area under the receiver operating characteristic curve ($AUC_{mean}$) across all folds.

Variables considered as potential risk factors in the voxel-level models included D, D weighted by a constant RBE (1.0 for photons and 1.1 for protons; $D_{constRBE}$), D multiplied by $LET_d$, and the PVR as a binary variable (1 for voxels within the PVR and 0 otherwise). In the combined proton-photon cohort, proton treatment was additionally evaluated both as separate risk factor and through interaction terms with other variables (e.g. $D{\cdot}LET_d{\cdot}$proton treatment). Voxels located within RICE were labeled as affected (responder, value = 1), whereas voxels outside RICE were labeled as non-affected (non-responder, value = 0). Voxels within the gross tumor volume and within cerebrospinal fluid were excluded for all patients. In addition, a dose threshold of 1 Gy was applied, and voxels receiving doses below this threshold were excluded. Because RICE may increase in volume over time, spatial information on lesion onset may be obscured. To account for this, volume-based exclusion criteria were applied in accordance with Bahn et al. [36] and Eulitz et al. [37]. Specifically, analyses were performed after excluding the largest 30—70% of RICE based on RICE volume. The applied volume thresholds and the corresponding numbers and characteristics of RICE before and after exclusion are provided in Supplementary Table S5 and Supplementary Figure S1. A variable RBE model was derived from the combined photon-proton voxel-level NTCP model that incorporated D, $D{\cdot}LET_d{\cdot}$proton treatment, and PVR. The derivation was based on the definition of RBE as the ratio of isoeffective doses for photon and proton radiation (see Supplementary Material).

In addition, we assessed whether patient-level risk in the photon, proton and combined photon-proton cohorts could be predicted using dose-volume histogram (DVH) parameters. The volume $V_x$ receiving more than x Gy(RBE) was evaluated for x values from 10 to 60 Gy(RBE) in 5 Gy(RBE) increments. The minimum dose $D_x$ to the volume x (in ml) receiving the highest dose was assessed for volumes ranging from 1 to 15 ml in 1 ml increments. All DVH parameters were calculated for the whole brain, the brain excluding the CTV, and the PVR, assuming either constant RBE values (photons: 1.0; protons: 1.1) or a variable proton RBE based on the formulation derived from voxel-level RICE risk modelling. In addition to DVH parameters, several clinical parameters were assessed as potential risk factors for RICE, including age at treatment initiation, sex, WHO tumor grade, tumor histology, tumor resection, prescribed total dose, CTV volume, treatment modality (photons or protons), and treatment technique (3D-CRT or IMRT for photons; DS or PBS for protons).

## Results

In total, 128 RICE were identified in 64 of 152 patients (42%; Table 1). RICE occurred in 32% of patients in the photon cohort and in 49% in the proton cohort. Within the proton cohort, RICE incidence was comparable between delivery techniques (DS: 51%; PBS: 46%). The median time from the start of radiotherapy to RICE detection was 14 months (range: 4–47) and did not differ significantly between modalities (proton: 14 months [DS: 14; PBS: 13] vs. photons: 16 months; $p = 0.28$; Table S6). The median number of RICE per affected patient was one in both photon and proton cohorts (DS: 2; PBS: 1), with ranges of 1-10 for photon- and proton-DS-treated patients and 1-5 for proton-PBS-treated patients. At first detection, RICE volume was significantly larger following proton PBS (median: 512 $mm^3$) compared to photons (86 $mm^3$; $p = 0.0016$) and proton DS (68 $mm^3$; $p = 0.0012$). Spatial RICE characteristics were consistent across treatment modalities, with median minimum distances of 0 mm between the RICE and CTV borders and 1 mm to the ventricles. In the photon cohort, 71% of RICE were located at least partly within the PVR (29/41); in the proton cohort, 81% (70/87; DS: 79% [53/67], PBS: 85% [17/20]). RICE volume, time to onset, and spatial relationships are illustrated in Figure 1.

Dose–volume and $LET_d$–volume histogram parameters within RICE differed significantly between treatment modalities and proton delivery techniques (Figure 2; Supplementary Table S6). Median near-minimum and mean $D_{constRBE}$ values were higher for photons than for protons, with the largest difference observed for the near-minimum dose ($D_{constRBE,98}$: 58 Gy(RBE) [12–62] vs. 53 Gy(RBE) [0–62]; $p = 0.013$). Within the proton cohort, near-minimum and mean $D_{constRBE}$ values were higher for DS than for PBS (e.g., $D_{constRBE,98}$: 55 Gy(RBE) [0–62] vs. 49 Gy(RBE) [0–61]; $p = 0.0049$). Mean and near-maximum $LET_d$ values were comparable between DS and PBS, whereas near-minimum $LET_d$ differed significantly. In addition, $D\cdot LET_d$–volume histogram parameters were consistently higher for DS than for PBS (e.g., $(D\cdot LET_d)_{98}$: 164 Gy·keV/µm [0–226] vs. 119 Gy·keV/µm [0–163]; $p<0.001$).

Several clinical parameters differed significantly between patients with and without RICE (Table 1; Supplementary Tables S1-4). In the combined photon-proton cohort, patients with RICE had larger CTV volumes and received higher prescribed total doses. RICE occurred in 9% of patients with meningioma and 65% of those with glioblastoma. Similarly, RICE was observed in 6% of patients with WHO grade I tumors compared to 63% with WHO grade IV tumors. Across cohorts, factors associated with RICE included larger CTV volume, glioblastoma histology, WHO tumor grade IV, increasing age, higher prescribed total dose, and surgery, with varying relevance between cohorts (Supplementary Table S7).

In univariate voxel-level analysis, D showed the highest predictive performance in the photon cohort ($AUC_{mean} = 0.86$; Table 2). In contrast, $D\cdot LET_d$ was the strongest predictor in all proton-containing cohorts ($AUC_{mean}$: 0.87 in the photon-proton cohort; 0.89 in the proton cohort [DS: 0.87; PBS: 0.93]), outperforming both D (photon-proton cohort: 0.71; proton: 0.81 [DS: 0.79; PBS: 0.81]) and $D_{constRBE}$ (photon-proton: 0.83). Across all cohorts, PVR was predictive of RICE ($AUC_{mean}$: 0.76 photon-proton; 0.63 photon; and 0.78 proton [DS: 0.77; PBS: 0.82]). In the combined cohort, proton treatment itself emerged as an additional risk factor ($AUC_{mean} = 0.68$). Multivariable voxel-level models yielded modest improvements in predictive performance. In the photon cohort, the combination of D and PVR achieved an $AUC_{mean}$ of 0.87 (Table 2). In the proton cohort, the model combining D, $D{\cdot}LET_d$, and PVR reached an $AUC_{mean}$ of 0.90 (DS: 0.90; PBS: 0.92). In the combined photon-proton cohort, the model including D, $D{\cdot}LET_d{\cdot}$proton treatment, and PVR achieved an $AUC_{mean}$ of 0.89. Excluding $D{\cdot}LET_d$ consistently reduced model performance across all proton-containing cohorts (Table 2), underscoring the importance of LET-related effects for RICE risk modeling.

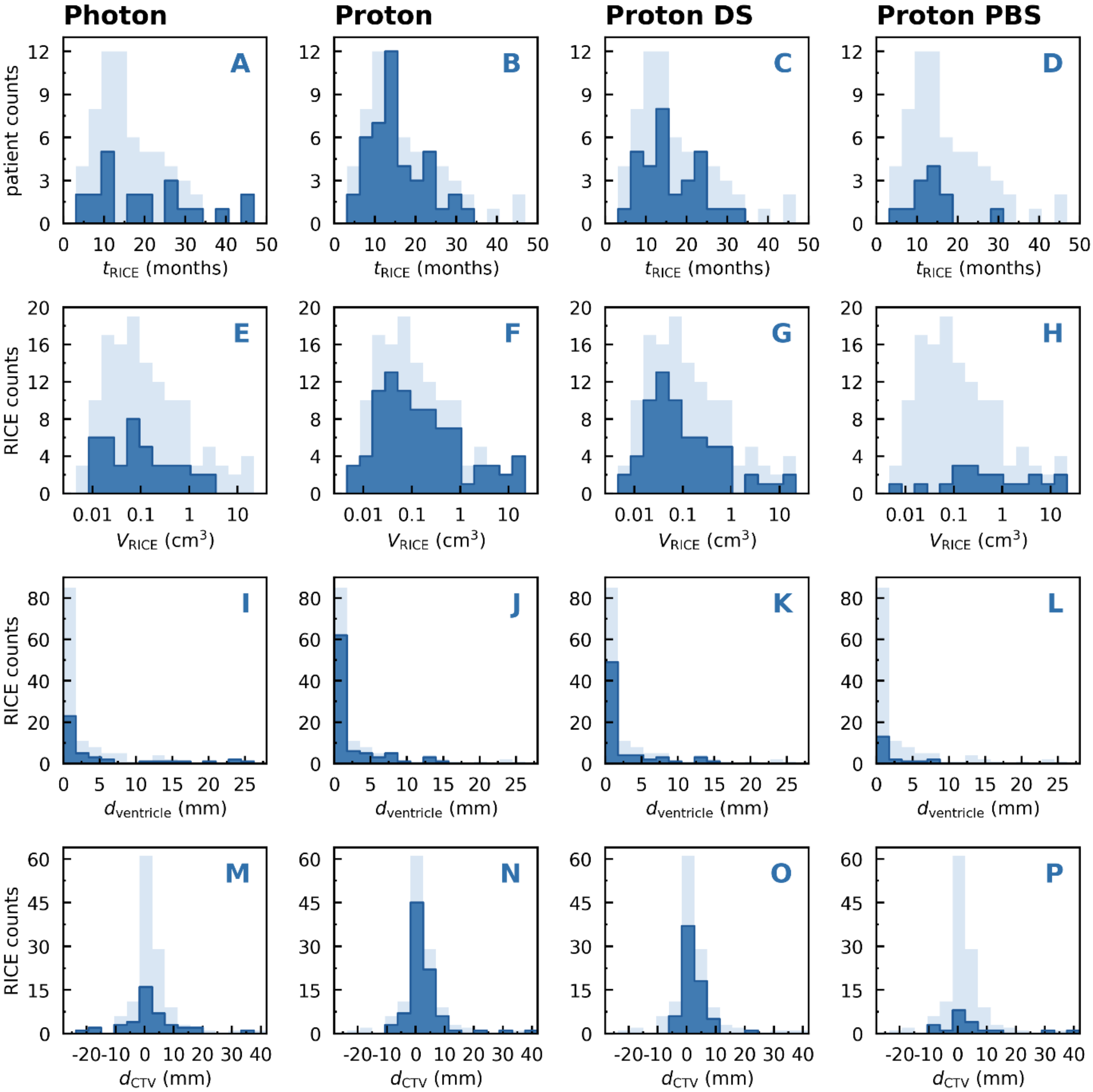

**FIGURE 1. Characteristics of radiation-induced contrast enhancement** (RICE). **(A-D)** Time from the start of radio(chemo)therapy to first RICE detection ($t_{RICE}$), **(E-H)** RICE volume at first detection ($V_{RICE}$), **(I-L)** distance from the RICE border to the cerebral ventricles ($d_{ventricle}$), and **(M-P)** distance from the RICE border to the clinical target volume ($d_{CTV}$). Distributions are shown for the photon (first column), proton (second column), proton double scattering (DS; third column), and proton pencil beam scanning (PBS; fourth column) cohorts (dark blue). The combined photon–proton cohort is shown in light blue. Distances were measured between the respective region-of-interest boundaries. Negative values of $d_{CTV}$ indicate that RICE is located within the CTV contour.

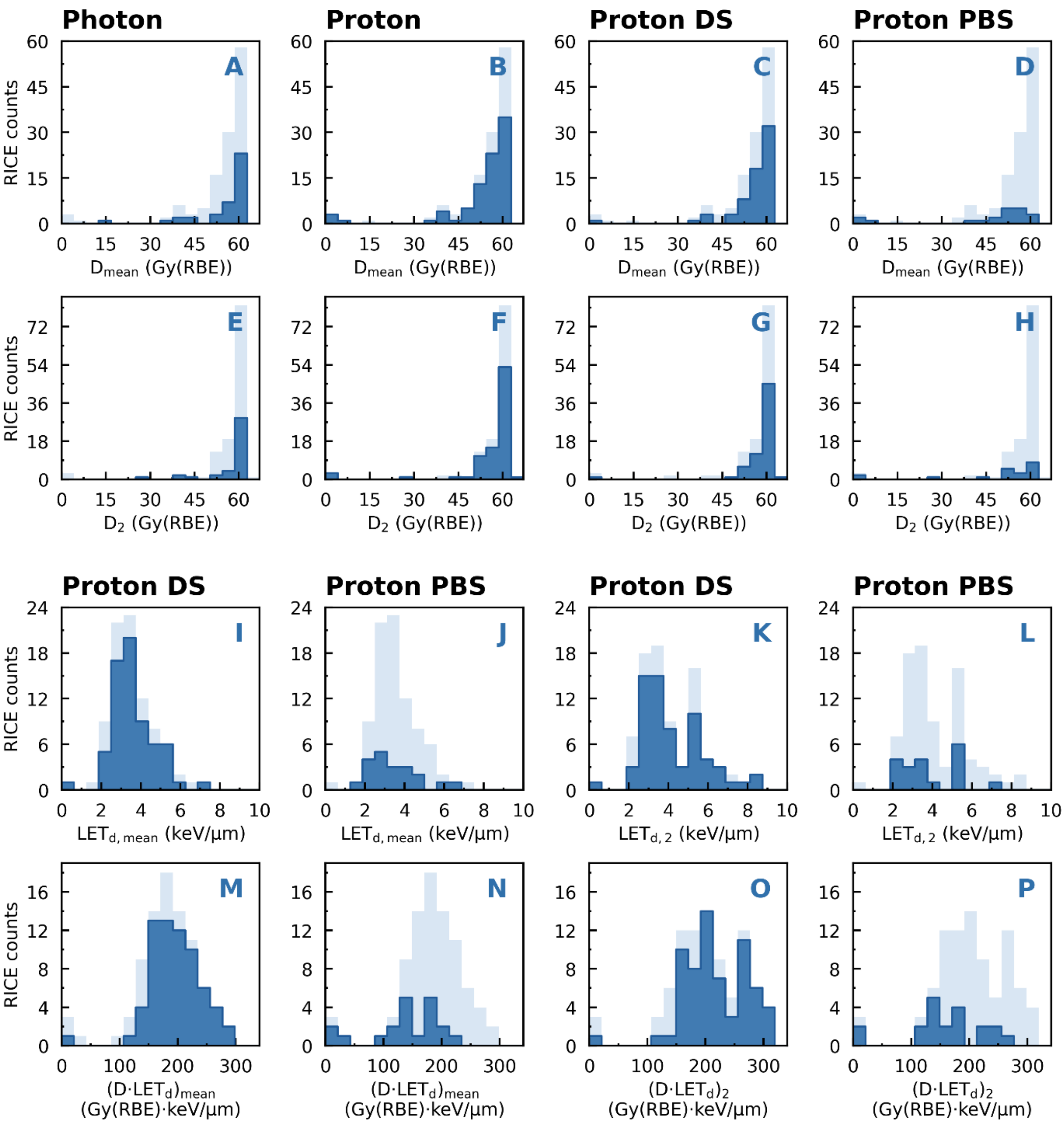

**FIGURE 2. Dose- and LET$_d$-volume histogram parameters of radiation induced contrast enhancement** (RICE). **(A-D)** $D_{mean}$, **(E-H)** $D_2$, **(I-J)** $LET_{d,mean}$, **(K-L)** $LET_{d,2}$, **(M-N)** $(D{\cdot}LET_d)_{mean}$, and **(O-P)** $(D{\cdot}LET_d)_2$. Distributions are shown for the photon (A,E), proton (B,F), proton double scattering (DS; C, G, I, M, K, O), and proton pencil beam scanning (PBS; D, H, J, L, N, P) cohorts (dark blue). The combined photon–proton cohort (A-H) or the proton cohort (I-P) is shown in light blue. *Abbreviations:* D, constant RBE-weighted dose; $D_{mean}$, mean D within the RICE volume; $D_2$, D in 2% of the RICE volume; $LET_d$, dose-weighted linear energy transfer; $LET_{d,mean}$, mean $LET_d$ in the RICE volume; $LET_{d,2}$, $LET_d$ in 2% of the RICE volume; $D{\cdot}LET_d$, product of D and $LET_d$; $(D{\cdot}LET_d)_{mean}$, mean $D{\cdot}LET_d$ in RICE volume; $(D{\cdot}LET_d)_2$, $D{\cdot}LET_d$ in 2% of the RICE volume; RBE, relative biological effectiveness.

Based on the combined photon-proton multivariable voxel-level model, a variable RBE model of the form $RBE = 1 + m \cdot LET_d$ was derived, yielding a slope parameter of $m = 0.10$ µm/keV. By comparison, an RBE model derived from the proton cohort alone, assuming a photon $LET_d$ of 0 keV/μm, resulted in a higher slope (m = 0.16 μm/keV; [DS: 0.15; PBS: 0.30]). Model performance metrics and parameter estimates are summarized in Table 2. Figure 3 illustrates voxel-wise predicted RICE risk, contributing factors, and derived RBE in an example patient. At the patient level, the best-performing RICE risk models included $D_{2ml}$(PVR) in the combined cohort ($AUC_{mean}$ = 0.78 [95% confidence interval (CI): [0.67-0.89]), $V_{10Gy(RBE)}$(brain) in the photon cohort (0.75 [0.56-0.93]), and $D_{2ml}$(brain) in the proton cohort (0.83 [0.69-0.97]). For cohorts including proton-treated patients, the best-performing models using constant-RBE-weighted dose yielded slightly lower AUC values (photon–proton cohort: 0.65 [0.52–0.78] vs. 0.78 [0.67–0.89]; proton cohort: 0.81 [0.67–0.96] vs. 0.83 [0.69–0.97]; Supplementary Table S8) and tended to show poorer calibration (Supplementary Figure S2). AUC values, regression coefficients, and calibration plots for the highest-performing DVH-based patient-level models derived from the brain, brain minus CTV, and PVR are provided in Supplementary Table S8 and Supplementary Figure S2.
For proton-containing cohorts, these models were based on the variable RBE derived from voxel-level modelling. The discriminatory performance of both voxel- and patient-level models is illustrated by receiver operating characteristic curves shown in Figure 4.

## Discussion

This study addresses two central questions in brain tumor radiotherapy: whether the clinical assumption of a constant proton RBE of 1.1 overlooks clinically relevant LET-dependent effects, and whether specific brain subvolumes, particularly the periventricular region, exhibit increased radiosensitivity. We analyzed radiation-induced contrast enhancement as a spatially resolved imaging endpoint in a combined cohort of proton- and photon-treated patients. Our results identified proximity to the cerebral ventricles and dose as consistent risk factors across radiation modalities. LET emerged as an additional risk factor in proton therapy, indicating increased biological effectiveness.
These findings extend previous RICE risk modeling studies that developed [32,34,36,37] and validated [42] predictive models in proton-only cohorts. However, these analyses precluded separating proton-specific effects from dose-response relationships also present in photon therapy. It remained unclear whether the observed periventricular radiosensitivity was modality-specific, and assessing RBE effects required assumptions regarding the reference dose-response relationship. By jointly analyzing photon- and proton-treated patients, the present study enables a direct comparison of dose-response relationships across modalities.
This combined analysis yields several insights. First, RICE occurs after both proton and photon therapy, indicating a general radiation response rather than a proton-specific phenomenon. Second, risk factors are largely shared across modalities, with increasing risk with dose and spatial clustering in the periventricular region. Notably, dose showed a comparable effect in both cohorts, as reflected by similar model-derived odds ratios, whereas dose-LET interaction emerged as an additional risk factor in proton therapy. These findings support a common underlying dose-response relationship, with LET contributing an additional effect related to beam quality (ionization density). Third, the presence of an elevated RICE risk in the periventricular region in both proton- and photon-treated patients suggests intrinsic rather than

**Table 2. Voxel-level modeling results.** Results were obtained using repeated 3-fold cross-validation with 333 repetitions. Regression coefficients $\beta_i$ are reported in units of $Gy^{-1}$ for the variable D, $Gy(RBE)^{-1}$ for the variable $D_{constRBE}$, $Gy(RBE)^{-1}$ µm/keV for the variables $D\cdot LET_d$ and $D\cdot LET_d\cdot$ proton treatment, and are dimensionless for all other variables. The parameter m is reported in µm/keV.

**Univariate modeling**

| **Variables** | **Photon-proton AUC** | **(95% CI)** | **Photon AUC** | **(95% CI)** | **Proton AUC** | **(95% CI)** | **Proton DS AUC** | **(95% CI)** | **Proton PBS AUC** | **(95% CI)** |
|---|---|---|---|---|---|---|---|---|---|---|
| D | 0.71 | (0.24-1.00) | 0.86 | (0.85-0.88) | 0.81 | (0.65-0.97) | 0.79 | (0.68-0.90) | 0.81 | (0.43-1.00) |
| $D_{constRBE}$ | 0.83 | (0.59-1.00) | 0.86 | (0.85-0.88) | 0.81 | (0.65-0.97) | 0.79 | (0.68-0.90) | 0.81 | (0.43-1.00) |
| $D\cdot LET_d$ | 0.87 | (0.86-0.88) | - | - | 0.89 | (0.89-0.90) | 0.87 | (0.86-0.88) | 0.93 | (0.77-1.00) |
| PVR | 0.76 | (0.75-0.77) | 0.63 | (0.59-0.67) | 0.78 | (0.77-0.79) | 0.77 | (0.76-0.79) | 0.82 | (0.79-0.84) |
| proton treatment | 0.68 | (0.67-0.69) | - | - | - | - | - | - | - | - |

**Multivariate modeling**

| **Variables** | **Cohort** | **AUC** | **(95% CI)** | **$-\beta_0$** | (95% CI) | **$\beta_1$** | (95% CI) | **$\beta_2$** | (95% CI) | **$\beta_3$** | (95% CI) | **$m=\beta_2/\beta_1$** |
|---|---|---|---|---|---|---|---|---|---|---|---|---|
| D, PVR | Photon-proton | 0.71 | (0.23-1.00) | 14.53 | (14.38-14.68) | 0.088 | (0.085-0.091) | 3.09 | (3.04-3.14) | | | |
| D, $D\cdot LET_d\cdot$proton treatment, PVR | Photon-proton | 0.89 | (0.82-0.97) | 18.93 | (18.60-19.25) | 0.15 | (0.14-0.15) | 0.015 | (0.015-0.016) | 2.91 | (2.85-2.96) | 0.10 |
| D, PVR | Photon | 0.87 | (0.85-0.89) | 18.58 | (17.55-19.61) | 0.12 | (0.10-0.14) | 2.00 | (1.73-2.27) | | | |
| D, PVR | Proton | 0.80 | (0.57-1.00) | 15.71 | (15.43-15.99) | 0.13 | (0.12-0.13) | 3.34 | (3.28-3.40) | | | |
| D, $D\cdot LET_d$, PVR | Proton | 0.90 | (0.85-0.96) | 19.90 | (19.49-20.31) | 0.14 | (0.13-0.15) | 0.023 | (0.022-0.024) | 2.92 | (2.85-2.98) | 0.16 |
| D, PVR | Proton DS | 0.85 | (0.66-1.00) | 15.27 | (14.94-15.60) | 0.12 | (0.11-0.13) | 3.31 | (3.24-3.38) | | | |
| D, $D\cdot LET_d$, PVR | Proton DS | 0.90 | (0.84-0.96) | 20.85 | (20.33-21.36) | 0.16 | (0.15-0.16) | 0.024 | (0.023-0.025) | 2.84 | (2.76-2.91) | 0.15 |
| D, PVR | Proton PBS | 0.80 | (0.35-1.00) | 16.00 | (15.45-16.56) | 0.12 | (0.11-0.13) | 3.57 | (3.41-3.74) | | | |
| D, $D\cdot LET_d$, PVR | Proton PBS | 0.92 | (0.81-1.00) | 21.30 | (20.32-22.28) | 0.13 | (0.11-0.14) | 0.038 | (0.035-0.041) | 3.18 | (3.01-3.35) | 0.30 |

*Abbreviations:* AUC, mean area under the receiver operating characteristic curve; CI, confidence interval; D, physical dose; $D_{constRBE}$, D weighted by a constant RBE of 1.1 for proton therapy and 1.0 for photon patients; DS, double scattering; $LET_d$, dose-weighted linear energy transfer; m, RBE model parameter (see Supplementary Material); PBS, pencil beam scanning; PVR, periventricular region; RBE, relative biological effectiveness.

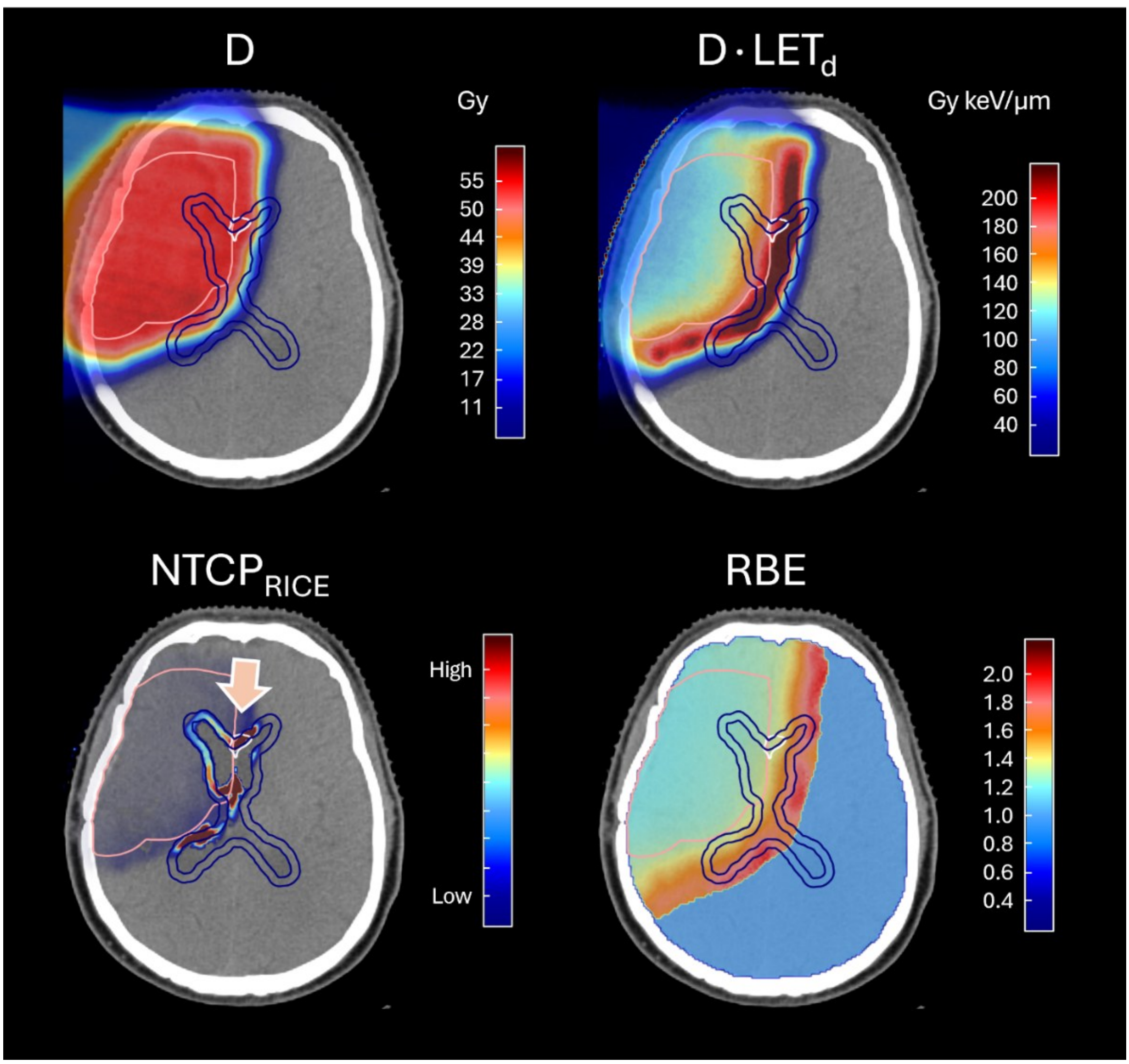

**FIGURE 3. Prediction of radiation-induced contrast enhancement (RICE) risk in a proton-treated patient.** Shown are distributions of the physical dose (D), D multiplied by dose-averaged LET ($D{\cdot}LET_d$), predicted probability of RICE ($NTCP_{RICE}$), and RBE derived from $NTCP_{RICE}$. Contours denote the clinical target volume (light red), RICE (white), and periventricular region (dark blue). A RICE developed in an area of elevated predicted RICE risk (arrow).

modality-specific radiosensitivity. The effect was more pronounced in the proton cohort, as reflected by higher odds ratios and AUC values. Finally, by comparing voxel-wise risks between modalities, we derive a clinically grounded RBE model for RICE that directly reflects the definition of RBE as the ratio of photon to proton dose producing the same biological effect. Several findings from this study support the clinical relevance of proton RBE variability. At the voxel level, models incorporating dose multiplied by $LET_d$ improved predictive performance compared with dose-only models and those assuming a constant RBE. Excluding LET-related information resulted in a systematic decline in model performance, suggesting that LET modifies the biological effectiveness of absorbed dose, consistent with established radiobiological evidence [5,51]. Based on these findings, we derived a voxel-level NTCP model yielding a variable RBE formulation of the form $RBE = 1 + m \cdot LET_d$, with $m = 0.10$ μm/keV. This estimate lies within the range reported in proton-only analyses (0.10–0.15 μm/keV) [34,36,37], but is lower than the values from our proton-only analyses, likely reflecting methodological differences. While proton-only approaches rely on implicit reference assumptions, the present analysis incorporates photon response as an explicit clinical baseline. At the patient level, RICE risk models based on DVH parameters calculated using the derived variable proton RBE outperformed models assuming a constant RBE, further supporting its clinical relevance. Moreover, under the assumption of a proton RBE of 1.1, RBE-weighted doses in RICE tended to be lower in proton than in photon patients, potentially reflecting an increased RBE. This is supported by clinical observations of higher rates of post-radiogenic contrast-enhancing lesions following proton compared with photon radiotherapy in glioma patients (47.5% [47/99] vs. 15.6% [22/141]; $p < 0.0001$) [41], suggesting that an underestimation of proton RBE may contribute to an increased risk.

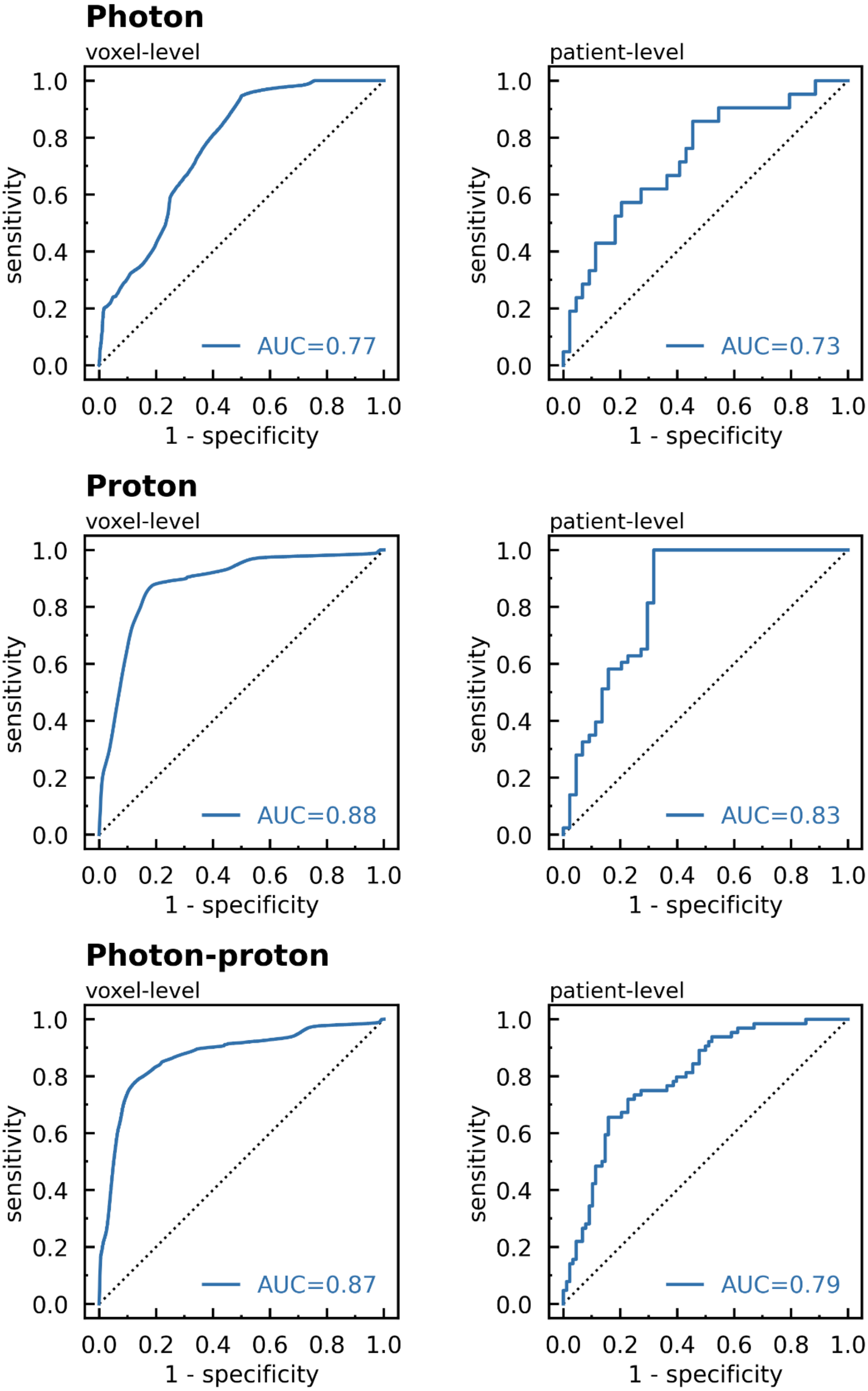

**FIGURE 4. Discriminatory performance of RICE risk prediction models.** Receiver operating characteristic curves for voxel-level (left) and patient-level (right) models are shown for the best-performing models in the photon (top), proton (middle), and combined photon–proton (bottom) cohorts. *Abbreviations*: AUC, area under the receiver operating characteristic curve.

Analysis of proton subcohorts revealed differences between delivery techniques. RICE volumes at first detection were larger in the PBS than in the DS cohort and were associated with lower $D{\cdot}LET_d$ values. The differences resulted in distinct NTCP model parameters and divergent RBE–$LET_d$ slopes (DS: 0.15 μm/keV; PBS: 0.30 μm/keV), indicating limited transferability of outcome models across delivery techniques. At the same time, combining DS and PBS data increases heterogeneity in dose and LET distributions, which may enhance detection of LET-dependent effects. In this context, historical DS data may still provide complementary insights relevant to current PBS treatments.

The findings of this study have several important clinical implications. First, the clinical assumption of a constant proton RBE of 1.1 may underestimate biological dose in regions of elevated LET, particularly near distal dose gradients, potentially increasing toxicity risk. Incorporating LET-dependent RBE variability into treatment planning may improve risk assessment and facilitate biologically optimized proton therapy. In this context, the interpretation of $D{\cdot}LET_d$ warrants careful consideration. Although $D{\cdot}LET_d$ emerged as a significant predictor in the RICE models, it is not directly interpretable in the same manner as conventional parameters such as physical dose or PVR. Instead, it should be viewed as a surrogate marker reflecting the LET-dependent component of biological effectiveness. This interpretation is consistent with the approximately linear relationship between biological effectiveness and LET assumed by both the RBE model developed in the present study and several established phenomenological RBE models [52]. Under this assumption, the RBE-weighted dose can be approximated as $D_{RBE} = D \cdot (1 + c \cdot LET_d) = D + c \cdot (D \cdot LET_d)$. Accordingly, $D{\cdot}LET_d$ alone neither represents biological dose nor implies biological equivalence across different dose–LET combinations. Rather, its predictive value supports the relevance of LET-dependent RBE modeling and underscores the importance of incorporating variable-RBE concepts into treatment optimization, rather than optimizing $D{\cdot}LET_d$ itself. Brain tumors represent a suitable clinical entity for investigating and optimizing proton RBE variability, owing to their relatively homogeneous tissue composition, limited anatomical variability, and high reproducibility in patient positioning [2,12,53,54]. Proton therapy in the brain typically involves low beam energies with short ranges, leading to steep distal LET gradients. Several organs at risk, such as the optic apparatus, are serially organized and exhibit low $\alpha/\beta$ ratios, conditions under which RBE increases may be more pronounced. Whether similar effects occur in other tumor entities should be addressed in future studies. Second, the consistent identification of the periventricular region as a radiosensitive structure supports its consideration as an organ at risk across modalities and motivates further investigation of other brain substructures. Third, clinical factors such as larger CTV volume and higher WHO tumor grade were associated with increase RICE risk, suggesting susceptibility to radiation-induced side effects is influenced by both patient- and disease-specific factors. Finally, RICE represents a surrogate imaging endpoint and does not necessarily correspond to symptomatic toxicity. However, its voxel-wise spatial resolution enables local correlation of tissue response with dose, LET, and anatomical risk factors, making it a sensitive tool for detecting LET-dependent effects that may be obscured in patient-level toxicity endpoints [2,21,23]. Prospective studies are required to determine for which clinical endpoints proton RBE variability is relevant and how it can be incorporated into clinical workflows. These include clinical trials integrating dose, LET, and anatomical risk factors into treatment planning [55,56], such as the ongoing INDIGO trial (ClinicalTrials.gov Identifier: NCT05964569), as well as multicenter analyses across treatment modalities.

This study has several limitations. Its retrospective design precludes causal inference and necessitates prospective validation. Uncertainties in RICE detection, segmentation, spatial registration, and dosimetric analysis may affect voxel-wise precision. In addition, the photon and proton cohorts were not randomized or matched, limiting direct between-cohort comparisons. Furthermore, the voxel-level modeling approach did not explicitly account for correlations among voxels within individual patients, nor for interpatient variability. Finally, external validation in independent, multi-institutional cohorts is necessary to confirm the robustness and generalizability of the proposed models.

## Conclusions

This study demonstrates the clinical relevance of LET-dependent variability in proton RBE and identifies the periventricular region as a potentially radiosensitive structure across treatment modalities. By integrating proton and photon patient data, we enable a direct comparison of dose-response relationships not achievable in proton-only analyses and reveal localized LET-driven effects. These findings support a treatment paradigm in which proton RBE variability and regional brain vulnerability are explicitly accounted for to mitigate radiation-induced toxicity. Prospective validation in comparative photon–proton and multicenter settings will be essential to support clinical implementation.

## Declaration of generative AI and AI-assisted technologies in the writing process

During the preparation of this work, the authors used ChatGPT (OpenAI, version 5) to improve readability and language of the manuscript. After using this service, the authors carefully reviewed and edited the content as necessary and take full responsibility for the content of the publication.

## CRediT authorship contribution statement

**Martina Palkowitsch:** Writing – review & editing, Writing – original draft, Visualization, Validation, Software, Methodology, Investigation, Formal analysis, Data curation, Conceptualization. **Larissa S. Kilian:** Writing – review & editing, Software, Methodology, Investigation, Data curation. **Fabian Hennings:** Writing – review & editing, Software, Data curation. **Armin Lühr:** Writing – review & editing, Conceptualization. **Justus Thiem:** Writing – review & editing, Resources, Data curation. **Arne Grey:** Writing – review & editing, Resources, Data curation. **Rebecca Bütof:** Writing – review & editing, Resources, Data curation. **Annekatrin Seidlitz:** Writing – review & editing, Resources. **Esther G.C. Troost:** Writing – review & editing, Resources. **Mechthild Krause:** Writing – review & editing, Resources. **Steffen Löck:** Writing – review & editing, Supervision, Methodology, Conceptualization.

## Declaration of competing interest

The authors declare the following financial interests/personal relationships which may be considered as potential competing interests: For the present study, the authors received no external financial support, neither for the study design or materials used, nor for the collection, analysis, and interpretation of data, nor for the writing of the publication. OncoRay has a research collaboration with RaySearch Laboratories. Dr. Mechthild Krause received funding for her research projects by Merck KGaA (2014–2018 for preclinical study; 2018–2020 for clinical study). Dr. Mechthild Krause was involved in a publicly funded (German Federal Ministry of Education and Research) project with the companies Medipan (2019–2022), Attomol GmbH (2019–2022), GA Generic Assays GmbH (2019–2022), Gesellschaft für medizinische und wissenschaftliche genetische Analysen (2019–2022), Lipotype GmbH (2019–2022) and PolyAn GmbH (2019–2022). Dr. Krause confirms that, to the best of her knowledge, none of the above- mentioned funding sources were involved in the preparation of this paper. Dr. Troost serves as member of the Scientific Advisory Board of IBA. She confirms that, to the best of her knowledge, findings of this work are not related to this conflict of interest. All remaining authors have declared no conflicts of interest.

## Acknowledgements

We thank Christian Hahn, Jan Eulitz, Jona Bensberg, and Lauritz Klünder for their valuable support in answering our questions regarding the reimplementation of the in-house developed Monte-Carlo simulation, image registration, image segmentation and normal tissue complication probability modeling framework.

## Appendix A. Supplementary data

Supplementary data for this article are available at the end of this document.

# Supplement

## Combined photon–proton modeling of radiation-induced brain imaging changes supports variability in proton relative biological effectiveness and increased periventricular radiosensitivity

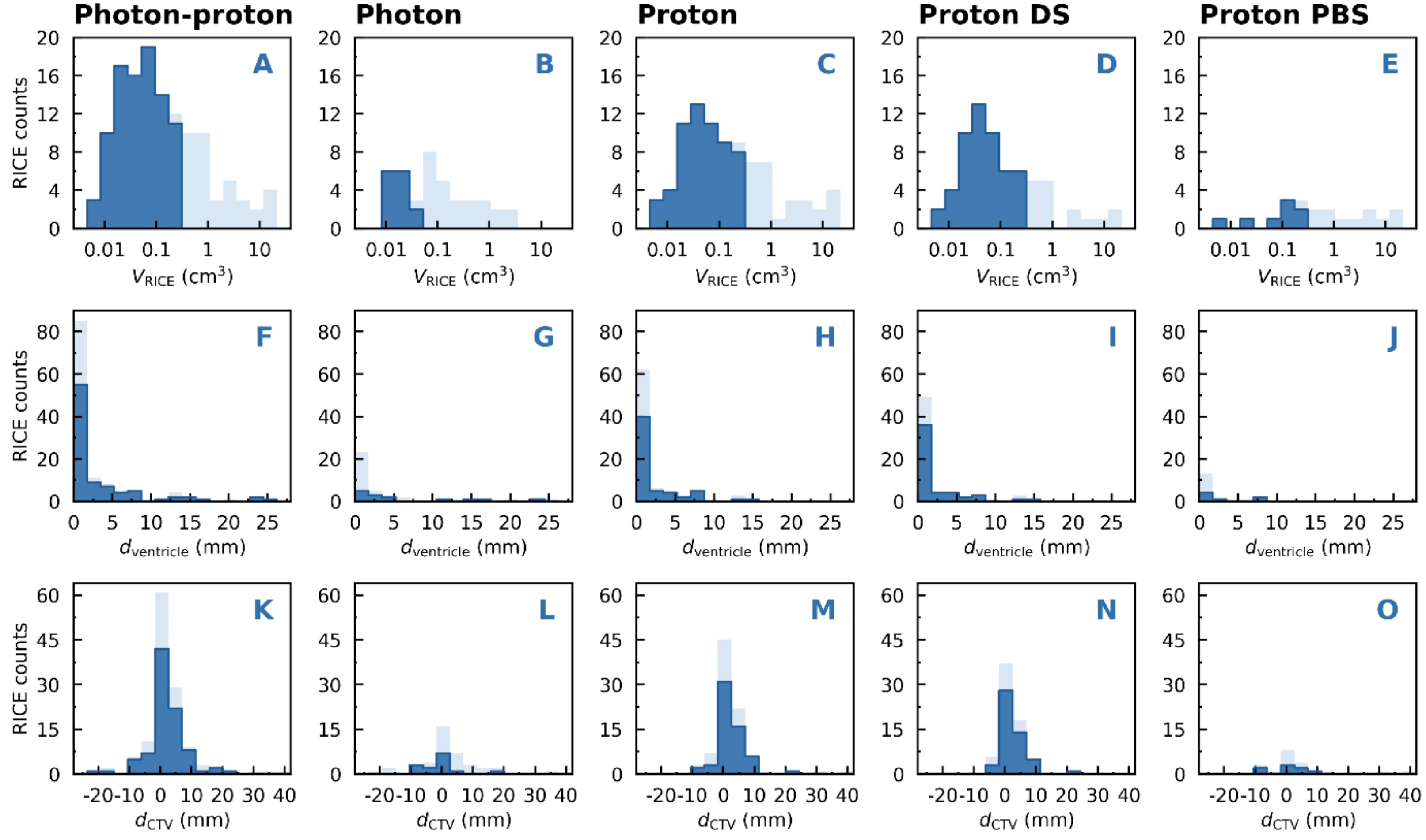

**FIGURE S1. Characteristics of radiation-induced contrast enhancement (RICE) included in the voxel-level analysis. (A-E)** RICE volume at first detection ($V_{RICE}$), **(F-J)** distance from the RICE border to the cerebral ventricles ($d_{ventricle}$), and **(K-O)** distance from the RICE border to the clinical target volume ($d_{CTV}$). Distributions are shown for the combined photon–proton cohort (first column), photon (second column), proton (third column), proton double scattering (DS; fourth column), and proton pencil beam scanning (PBS; fifth column) cohorts (dark blue). Dark blue denotes RICE included in the voxel-level analysis, while light blue denotes the full cohort, including RICE excluded from the voxel-level analysis. Distances were measured between the respective region-of-interest boundaries. Negative values of $d_{CTV}$ indicate that RICE is located within the CTV contour.

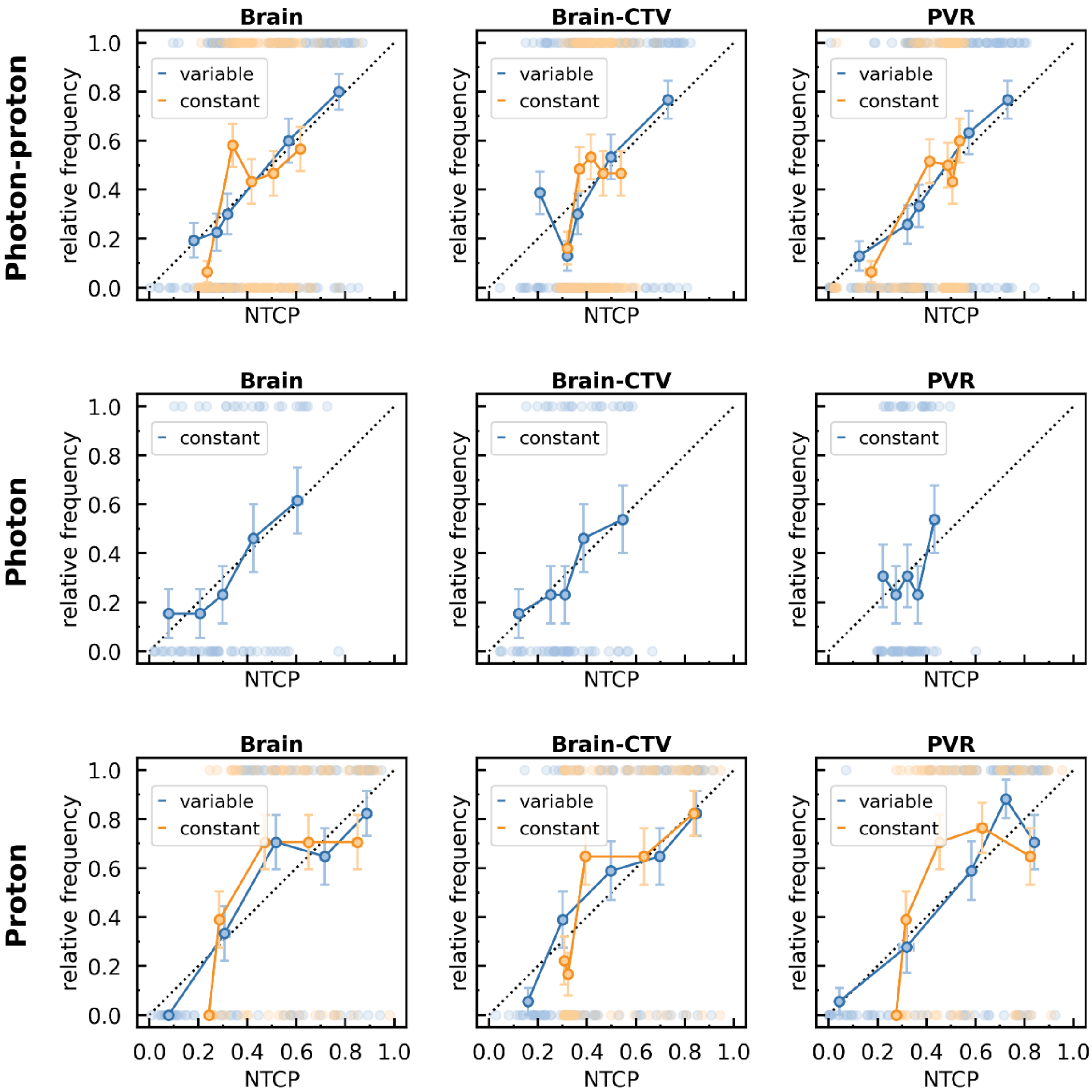

**FIGURE S2. Calibration plots for patient-level modeling.** Calibration plots are shown for the dose–volume histogram (DVH)-based patient-level models with the highest mean AUC in cross-validation for the combined photon–proton (first row), photon (second row), and proton cohorts (third row). Columns correspond to DVH parameters derived for the brain (first column), brain minus CTV (second column), and PVR (third column). For cohorts including proton-treated patients, results are shown for both constant- and variable-RBE-weighted doses (constant RBE: 1.0 for photons and 1.1 for protons, yellow; variable RBE: 1.0 for photons and $\mathrm{RBE} = 1 + m \cdot \mathrm{LET}_d$ for protons, with $m = 0.10$ µm/keV derived from voxel-level analyses, blue). Semi-transparent points represent individual patient observations. Markers indicate the mean predicted NTCP and observed event frequency within equally sized bins, with SEM error bars. Solid lines connect adjacent bins, and the diagonal dotted line indicates perfect calibration.

*Abbreviations:* CTV, clinical target volume; NTCP, normal tissue complication probability; PVR, periventricular region.

**TABLE S1. Patient characteristics of the photon cohort.** Two-sided p-values were calculated to assess differences between patients with and without radiation-induced contrast enhancement (RICE). The $\chi^2$ test was used for categorical variables and the Mann–Whitney U test for continuous variables.

| **Variables** | **All patients** | | **Non-RICE patients** | | **RICE patients** | | |
|---|---|---|---|---|---|---|---|
| | n | (%) | n | (%) | n | (%) | |
| Number of patients | 65 | (100) | 44 | (68) | 21 | (32) | |
| Treatment period | 2012-2018 | | 2012-2018 | | 2012-2018 | | |
| | Median | (range) | Median | (range) | Median | (range) | p-value |
| Age (years) | 51 | (24-82) | 45 | (24-76) | 57 | (32-82) | 0.0071 |
| Clinical target volume (cm$^3$) | 228 | (47-569) | 210 | (47-440) | 253 | (125-569) | 0.023 |
| Prescribed dose (Gy (RBE)) | 60 | (54-60) | 60 | (54-60) | 60 | (54-60) | 0.45 |
| Dose per fraction (Gy(RBE)) | 2.0 | (1.8-2.0) | 2.0 | (1.8-2.0) | 2.0 | (2.0-2.0) | 0.51 |
| | n | (%) | n | (%) | n | (%) | p-value |
| Treatment technique | | | | | | | 0.48 |
| 3D-CRT | 32 | (49) | 23 | (72) | 9 | (28) | |
| IMRT | 33 | (51) | 21 | (64) | 12 | (36) | |
| Sex | | | | | | | 0.48 |
| Female | 32 | (49) | 23 | (72) | 9 | (28) | |
| Male | 33 | (51) | 21 | (64) | 12 | (36) | |
| Histology | | | | | | | < 0.001 |
| Astrocytoma | 12 | (18) | 12 | (100) | 0 | (0) | |
| Craniopharyngioma | 0 | (0) | 0 | (0) | 0 | (0) | |
| Ependymoma | 2 | (3) | 2 | (100) | 0 | (0) | |
| Glioblastoma | 34 | (52) | 15 | (44) | 19 | (56) | |
| Glioma | 1 | (2) | 1 | (100) | 0 | (0) | |
| Hemangiopericytoma | 0 | (0) | 0 | (0) | 0 | (0) | |
| Meningioma | 2 | (3) | 2 | (100) | 0 | (0) | |
| Oligoastrocytoma | 12 | (18) | 12 | (100) | 0 | (0) | |
| Oligodendroglioma | 2 | (3) | 0 | (0) | 2 | (100) | |
| Pituitary adenoma | 0 | (0) | 0 | (0) | 0 | (0) | |
| Unknown | 0 | (0) | 0 | (0) | 0 | (0) | |
| WHO tumor grade | | | | | | | < 0.001 |
| 1 | 0 | (0) | 0 | (0) | 0 | (0) | |
| 2 | 4 | (6) | 3 | (75) | 1 | (25) | |
| 3 | 27 | (42) | 26 | (96) | 1 | (4) | |
| 4 | 34 | (52) | 15 | (44) | 19 | (56) | |
| Tumor resection | | | | | | | 0.74 |
| Yes | 57 | (88) | 39 | (68) | 18 | (32) | |
| No | 8 | (12) | 5 | (62) | 3 | (38) | |

*Abbreviations:* 3D-CRT, three-dimensional conformal radiation therapy; IMRT, intensity-modulated radiotherapy; RBE, relative biological effectiveness; RICE, radiation-induced contrast enhancement.

**TABLE S2. Patient characteristics of the proton cohort.** Two-sided p-values were calculated to assess differences between patients with and without radiation-induced contrast enhancement (RICE). The $\chi^2$ test was used for categorical variables and the Mann–Whitney U test for continuous variables.

| **Variables** | **All patients** | | **Non-RICE patients** | | **RICE patients** | | |
|---|---|---|---|---|---|---|---|
| | n | (%) | n | (%) | n | (%) | |
| Number of patients | 87 | (100) | 44 | (51) | 43 | (49) | |
| Treatment period | 2014-2019 | | 2015-2019 | | 2014-2019 | | |
| | Median | (range) | Median | (range) | Median | (range) | p-value |
| Age (years) | 50 | (26-82) | 50 | (26-79) | 50 | (27-82) | 0.44 |
| Clinical target volume (cm$^3$) | 167 | (5-501) | 117 | (5-501) | 191 | (26-407) | < 0.001 |
| Prescribed dose (Gy (RBE)) | 60 | (54-60) | 54 | (54-60) | 60 | (54-60) | < 0.001 |
| Dose per fraction (Gy(RBE)) | 2.0 | (1.8-2.0) | 2.0 | (1.8-2.0) | 2.0 | (2.0-2.0) | 0.16 |
| | n | (%) | n | (%) | n | (%) | p-value |
| Treatment technique | | | | | | | 0.69 |
| Double scattering | 61 | (70) | 30 | (49) | 31 | (51) | |
| Pencil beam scanning | 26 | (30) | 14 | (54) | 12 | (46) | |
| Sex | | | | | | | 0.11 |
| Female | 48 | (55) | 28 | (58) | 20 | (42) | |
| Male | 39 | (45) | 16 | (41) | 23 | (59) | |
| Histology | | | | | | | < 0.001 |
| Astrocytoma | 22 | (25) | 9 | (41) | 13 | (59) | |
| Craniopharyngioma | 1 | (1) | 1 | (100) | 0 | (0) | |
| Ependymoma | 2 | (2) | 2 | (100) | 0 | (0) | |
| Glioblastoma | 17 | (20) | 3 | (18) | 14 | (82) | |
| Glioma | 1 | (1) | 1 | (100) | 0 | (0) | |
| Hemangiopericytoma | 1 | (1) | 0 | (0) | 1 | (100) | |
| Meningioma | 21 | (24) | 19 | (90) | 2 | (10) | |
| Oligoastrocytoma | 4 | (5) | 1 | (25) | 3 | (75) | |
| Oligodendroglioma | 15 | (17) | 5 | (33) | 10 | (67) | |
| Pituitary adenoma | 1 | (1) | 1 | (100) | 0 | (0) | |
| Unkown | 2 | (2) | 2 | (100) | 0 | (0) | |
| WHO tumor grade | | | | | | | < 0.001 |
| 1 | 18 | (21) | 17 | (94) | 1 | (6) | |
| 2 | 17 | (20) | 9 | (53) | 8 | (47) | |
| 3 | 31 | (36) | 12 | (39) | 19 | (61) | |
| 4 | 18 | (21) | 4 | (22) | 14 | (78) | |
| n.g. | 3 | (3) | 2 | (67) | 1 | (33) | |
| Tumor resection | | | | | | | 0.0038 |
| Yes | 65 | (75) | 27 | (42) | 38 | (58) | |
| No | 22 | (25) | 17 | (77) | 5 | (23) | |

*Abbreviations:* RBE, relative biological effectiveness; RICE, radiation-induced contrast enhancement.

**TABLE S3. Patient characteristics of the proton DS cohort.** Two-sided p-values were calculated to assess differences between patients with and without radiation-induced contrast enhancement (RICE). The $\chi^2$ test was used for categorical variables and the Mann–Whitney U test for continuous variables.

| Variables | All patients | | Non-RICE patients | | RICE patients | | |
|---|---|---|---|---|---|---|---|
| | n | (%) | n | (%) | n | (%) | |
| Number of patients | 61 | (100) | 30 | (49) | 31 | (51) | |
| Treatment period | 2014-2018 | | 2015-2017 | | 2014-2018 | | |
| | Median | (range) | Median | (range) | Median | (range) | p-value |
| Age (years) | 46 | (26-79) | 47 | (26-79) | 41 | (27-72) | 0.49 |
| Clinical target volume (cm³) | 168 | (5-501) | 100 | (5-501) | 191 | (101-407) | 0.0032 |
| Prescribed dose (Gy (RBE)) | 60 | (54-60) | 54 | (54-60) | 60 | (54-60) | < 0.001 |
| Dose per fraction (Gy(RBE)) | 2.0 | (2.0-2.0) | 2.0 | (2.0-2.0) | 2.0 | (2.0-2.0) | |
| | n | (%) | n | (%) | n | (%) | p-value |
| Sex | | | | | | | 0.69 |
| Female | 33 | (54) | 17 | (52) | 16 | (48) | |
| Male | 28 | (46) | 13 | (46) | 15 | (54) | |
| Histology | | | | | | | 0.011 |
| Astrocytoma | 17 | (28) | 6 | (35) | 11 | (65) | |
| Craniopharyngioma | 1 | (2) | 1 | (100) | 0 | (0) | |
| Ependymoma | 1 | (2) | 1 | (100) | 0 | (0) | |
| Glioblastoma | 12 | (20) | 3 | (25) | 9 | (75) | |
| Glioma | 1 | (2) | 1 | (100) | 0 | (0) | |
| Hemangiopericytoma | 0 | (0) | 0 | (0) | 0 | (0) | |
| Meningioma | 13 | (21) | 12 | (92) | 1 | (8) | |
| Oligoastrocytoma | 4 | (6) | 1 | (25) | 3 | (75) | |
| Oligodendroglioma | 10 | (16) | 3 | (30) | 7 | (70) | |
| Pituitary adenoma | 1 | (2) | 1 | (100) | 0 | (0) | |
| Unknown | 1 | (2) | 1 | (100) | 0 | (0) | |
| WHO tumor grade | | | | | | | < 0.001 |
| 1 | 14 | (23) | 14 | (100) | 0 | (0) | |
| 2 | 9 | (15) | 1 | (11) | 8 | (89) | |
| 3 | 24 | (39) | 10 | (42) | 14 | (58) | |
| 4 | 13 | (21) | 4 | (31) | 9 | (69) | |
| n.g. | 1 | (2) | 1 | (100) | 0 | (0) | |
| Tumor resection | | | | | | | 0.0060 |
| Yes | 46 | (75) | 18 | (39) | 28 | (61) | |
| No | 15 | (25) | 12 | (80) | 3 | (20) | |

*Abbreviations:* DS, double scattering; n.g., not given; RBE, relative biological effectiveness; RICE, radiation-induced contrast enhancement.

**TABLE S4. Patient characteristics of the proton PBS cohort.** Two-sided p-values were calculated to assess differences between patients with and without radiation-induced contrast enhancement (RICE). The $\chi^2$ test was used for categorical variables and the Mann–Whitney U test for continuous variables.

| **Variables** | **All patients** | | **Non-RICE patients** | | **RICE patients** | | |
|---|---|---|---|---|---|---|---|
| | n | (%) | n | (%) | n | (%) | |
| Number of patients | 26 | (100) | 14 | (54) | 12 | (46) | |
| Treatment period | 2018-2019 | | 2018-2019 | | 2018-2019 | | |
| | Median | (range) | Median | (range) | Median | (range) | p-value |
| Age (years) | 58 | (31-82) | 54 | (40-72) | 60 | (31-82) | 0.64 |
| Clinical target volume ($cm^3$) | 154 | (26-315) | 141 | (36-210) | 200 | (26-315) | 0.11 |
| Prescribed dose (Gy (RBE)) | 60 | (54-60) | 54 | (54-60) | 60 | (60-60) | 0.0023 |
| Dose per fraction (Gy(RBE)) | 2.0 | (1.8-2.0) | 2.0 | (1.8-2.0) | 2.0 | (2.0-2.0) | 0.20 |
| | n | (%) | n | (%) | n | (%) | p-value |
| Sex | | | | | | | 0.020 |
| Female | 15 | (58) | 11 | (73) | 4 | (27) | |
| Male | 11 | (42) | 3 | (27) | 8 | (73) | |
| Histology | | | | | | | 0.046 |
| Astrocytoma | 5 | (19) | 3 | (60) | 2 | (40) | |
| Craniopharyngioma | 0 | (0) | 0 | (0) | 0 | (0) | |
| Ependymoma | 1 | (4) | 1 | (100) | 0 | (0) | |
| Glioblastoma | 5 | (19) | 0 | (0) | 5 | (100) | |
| Glioma | 0 | (0) | 0 | (0) | 0 | (0) | |
| Hemangiopericytoma | 1 | (4) | 0 | (0) | 1 | (100) | |
| Meningioma | 8 | (31) | 7 | (88) | 1 | (12) | |
| Oligoastrocytoma | 0 | (0) | 0 | (0) | 0 | (0) | |
| Oligodendroglioma | 5 | (19) | 2 | (40) | 3 | (60) | |
| Pituitary adenoma | 0 | (0) | 0 | (0) | 0 | (0) | |
| Unknown | 1 | (4) | 1 | (100) | 0 | (0) | |
| WHO tumor grade | | | | | | | 0.0043 |
| 1 | 4 | (15) | 3 | (75) | 1 | (25) | |
| 2 | 8 | (31) | 8 | (100) | 0 | (0) | |
| 3 | 7 | (27) | 2 | (29) | 5 | (71) | |
| 4 | 5 | (19) | 0 | (0) | 5 | (100) | |
| n.g. | 2 | (8) | 1 | (50) | 1 | (50) | |
| Tumor resection | | | | | | | 0.28 |
| Yes | 19 | (73) | 9 | (47) | 10 | (53) | |
| No | 7 | (27) | 5 | (71) | 2 | (29) | |

*Abbreviations:* n.g., not given; PBS, pencil beam scanning; RBE, relative biological effectiveness; RICE, radiation-induced contrast enhancement.

**TABLE S5. Exclusion of radiation-induced contrast enhancement (RICE) from the voxel-level analyses.** For the combined photon–proton, photon, proton, proton double-scattering (DS), and proton pencil-beam scanning (PBS) cohorts, the table reports the total number of RICE, the number included in the voxel-level analyses, the volume threshold used for exclusion, and the median, mean, minimum, and maximum RICE volumes. Volumes are reported in $mm^3$.

| **Cohort** | **n** | **(%)** | **RICE volume ($mm^3$)** | | | |
|---|---|---|---|---|---|---|
| | | | **Threshold** | **Median** | **Mean** | **(range)** |
| Photon-proton | 128 | (100) | - | 91 | 1064 | (5-21801) |
| Photon | 41 | (100) | - | 86 | 334 | (9-2971) |
| Proton | 87 | (100) | - | 102 | 1407 | (5-21801) |
| Proton DS | 67 | (100) | - | 68 | 835 | (5-15373) |
| Proton PBS | 20 | (100) | - | 512 | 3324 | (7-21801) |
| Photon-proton | 90 | (70) | 266 | 50 | 74 | (5-266) |
| Photon | 14 | (34) | 47 | 16 | 19 | (9-47) |
| Proton | 59 | (67) | 266 | 50 | 77 | (5-266) |
| Proton DS | 51 | (76) | 266 | 47 | 70 | (5-266) |
| Proton PBS | 8 | (40) | 266 | 107 | 120 | (7-265) |

**Table S6. RICE characteristics across cohorts.**

| Variables | Photon-proton | | Photon | | Proton | | Proton DS | | Proton PBS | |
|---|---|---|---|---|---|---|---|---|---|---|
| | Median | (range) | Median | (range) | Median | (range) | Median | (range) | Median | (range) |
| number of RICE per patient | 1 | (1-10) | 1 | (1-10) | 1 | (1-10) | 2 | (1-10) | 1 | (1-5) |
| time to RICE detection (months) | 14 | (4-47) | 16 | (4-47) | ,14 | (4-31) | 14 | (4-31) | 13 | (4-31) |
| RICE volume ($mm^3$) | 91 | (5-21801) | 86 | (9-2971) | 102 | (5-21801) | 68 | (5-15373) | 512 | (7-21801) |
| $d_{vent}$ (mm) | 1 | (0-26) | 1 | (0-26) | 1 | (0-15) | 1 | (0-15) | 0 | (0-8) |
| $d_{CTV}$ (mm) | 0 | (-22-39) | 0 | (-22-34) | 0 | (-9-39) | 0 | (-6-24) | 0 | (-9-39) |
| $D_{constRBE,2}$ (Gy(RBE)) | 60 | (0-63) | 60 | (28-63) | 60 | (0-63) | 60 | (0-63) | 57 | (0-62) |
| $D_{constRBE,98}$ (Gy(RBE)) | 55 | (0-62) | 58 | (12-62) | 53 | (0-62) | 55 | (0-62) | 49 | (0-61) |
| $D_{constRBE,mean}$ (Gy(RBE)) | 58 | (0-62) | 59 | (16-62) | 57 | (0-62) | 59 | (0-62) | 53 | (0-62) |
| $LET_{d,2}$ (kev/μm) | 3 | (0-12) | 0.3 | (0.3-0.3) | 4 | (0-12) | 4 | (0-9) | 4 | (2-12) |
| $LET_{d,98}$ (kev/μm) | 3 | (0-6) | 0.3 | (0.3-0.3) | 3 | (0-6) | 3 | (0-6) | 3 | (0-4) |
| $LET_{d,mean}$ (kev/μm) | 3 | (0-7) | 0.3 | (0.3-0.3) | 3 | (0-7) | 3 | (0-7) | 3 | (2-7) |
| $(D \cdot LET_d)_2$ (Gy·kev/μm) | 153 | (0-290) | 19 | (9-19) | 182 | (0-290) | 192 | (0-290) | 141 | (2-233) |
| $(D \cdot LET_d)_{98}$ (Gy·kev/μm) | 129 | (0-226) | 18 | (4-19) | 151 | (0-226) | 164 | (0-226) | 119 | (0-163) |
| $(D \cdot LET_d)_{mean}$ (Gy·kev/μm) | 143 | (0-256) | 18 | (5-19) | 168 | (0-256) | 176 | (0-256) | 125 | (1-198) |

*Abbreviations:* CTV, clinical target volume; D, physical dose; $D_{constRBE}$, D weighted by a constant RBE of 1.1 for proton therapy and 1.0 for photon patients; $D_{constRBE,mean}$, mean $D_{constRBE}$ within the RICE volume; $D_{constRBE,x}$, $D_{constRBE}$ in x% of the RICE volume; $d_{ctv}$, distance between RICE and CTV; $D \cdot LET_d$, product of D and $LET_d$; $(D \cdot LET_d)_{mean}$, mean $D \cdot LET_d$ in RICE volume; $(D \cdot LET_d)_x$, $D \cdot LET_d$ in x% of the RICE volume; DS, double scattering; $d_{vent}$ distance between ventricles and RICE; $LET_d$, dose-weighted linear energy transfer; $LET_{mean}$, mean $LET_d$ in the RICE volume; $LET_{d,x}$, $LET_d$ in x% of the RICE volume; PBS, pencil beam scanning; RBE, relative biological effectiveness; RICE, radiation-induced contrast enhancement.

**TABLE S7. Patient-level modeling results for clinical parameters.** Results were obtained using repeated 3-fold cross-validation with 333 repetitions.

| | Photon-proton | | Photon | | Proton | | Proton DS | | Proton PBS | |
|---|---|---|---|---|---|---|---|---|---|---|
| **Variables** | **AUC** | **(95% CI)** | **AUC** | **(95% CI)** | **AUC** | **(95% CI)** | **AUC** | **(95% CI)** | **AUC** | **(95% CI)** |
| Treatment modality | 0.59 | (0.47-0.70) | - | - | - | - | - | - | - | - |
| Age | 0.53 | (0.36-0.70) | 0.71 | (0.52-0.90) | 0.50 | (0.29-0.70) | 0.52 | (0.28-0.76) | 0.40 | (0.04-0.75) |
| Clinical target volume | 0.71 | (0.59-0.83) | 0.78 | (0.60-0.96) | 0.78 | (0.64-0.93) | 0.77 | (0.58-0.95) | 0.79 | (0.47-1.00) |
| Prescribed dose | 0.62 | (0.53-0.71) | 0.49 | (0.38-0.60) | 0.76 | (0.64-0.88) | 0.76 | (0.60-0.91) | 0.79 | (0.58-0.99) |
| Treatment technique | | | | | | | | | | |
| 3D-CRT | 0.56 | (0.46-0.66) | 0.49 | (0.28-0.70) | - | - | - | - | - | - |
| IMRT | 0.49 | (0.39-0.60) | 0.49 | (0.28-0.70) | - | - | - | - | - | - |
| Double scattering | 0.57 | (0.45-0.69) | - | - | 0.46 | (0.34-0.58) | - | - | - | - |
| Pencil beam scanning | 0.47 | (0.39-0.56) | - | - | 0.46 | (0.34-0.58) | - | - | - | - |
| Sex | 0.56 | (0.43-0.68) | 0.49 | (0.28-0.70) | 0.58 | (0.42-0.74) | 0.44 | (0.28-0.61) | 0.72 | (0.45-1.00) |
| Histology | | | | | | | | | | |
| Astrocytoma | 0.48 | (0.39-0.58) | 0.64 | (0.54-0.73) | 0.52 | (0.37-0.68) | 0.56 | (0.37-0.75) | 0.42 | (0.23-0.61) |
| Craniopharyngioma | - | - | - | - | 0.50 | (0.50-0.50) | 0.50 | (0.50-0.50) | - | - |
| Ependymoma | 0.52 | (0.49-0.55) | 0.52 | (0.48-0.55) | 0.52 | (0.48-0.55) | 0.50 | (0.50-0.50) | 0.50 | (0.50-0.50) |
| Glioblastoma | 0.66 | (0.54-0.77) | 0.78 | (0.64-0.92) | 0.63 | (0.51-0.74) | 0.59 | (0.45-0.74) | 0.71 | (0.47-0.94) |
| Glioma | - | - | 0.50 | (0.50-0.50) | 0.50 | (0.50-0.50) | 0.50 | (0.50-0.50) | - | - |
| Hemangiopericytoma | - | - | - | - | 0.50 | (0.50-0.50) | - | - | 0.50 | (0.50-0.50) |
| Meningioma | 0.60 | (0.53-0.67) | 0.52 | (0.48-0.55) | 0.69 | (0.57-0.82) | 0.68 | (0.55-0.82) | 0.71 | (0.47-0.94) |
| Oligoastrocytoma | 0.55 | (0.49-0.62) | 0.64 | (0.54-0.73) | 0.51 | (0.44-0.58) | 0.51 | (0.41-0.62) | - | - |
| Oligodendroglioma | 0.56 | (0.48-0.65) | 0.53 | (0.46-0.61) | 0.55 | (0.42-0.68) | 0.55 | (0.39-0.70) | 0.46 | (0.20-0.72) |
| Pituitary adenoma | - | - | - | - | 0.50 | (0.50-0.50) | 0.50 | (0.50-0.50) | - | - |
| WHO tumor grade | | | | | | | | | | |
| 1 | 0.59 | (0.53-0.65) | - | - | 0.68 | (0.57-0.79) | 0.73 | (0.61-0.86) | 0.52 | (0.27-0.76) |
| 2 | 0.47 | (0.42-0.52) | 0.47 | (0.40-0.55) | 0.45 | (0.37-0.54) | 0.61 | (0.49-0.74) | 0.78 | (0.58-0.99) |
| 3 | 0.55 | (0.43-0.68) | 0.77 | (0.64-0.90) | 0.58 | (0.42-0.74) | 0.52 | (0.31-0.72) | 0.62 | (0.33-0.92) |
| 4 | 0.65 | (0.54-0.76) | 0.78 | (0.64-0.92) | 0.62 | (0.50-0.74) | 0.57 | (0.41-0.73) | 0.71 | (0.47-0.94) |
| Tumor resection | 0.56 | (0.48-0.65) | 0.46 | (0.35-0.56) | 0.64 | (0.51-0.76) | 0.65 | (0.50-0.80) | 0.56 | (0.29-0.84) |

*Abbreviations:* 3D-CRT, three-dimensional conformal radiation therapy; AUC, mean area under the receiver operating characteristic curve; CI, confidence interval; DS, double scattering; IMRT, intensity-modulated radiotherapy; PBS, pencil beam scanning.

**TABLE S8. Patient-level modeling results for dose–volume histogram parameters.** Results were obtained using repeated 3-fold cross-validation (333 repetitions). For each ROI (brain, brain–CTV, and PVR), the model with the highest mean AUC is shown for the photon–proton, photon, and proton cohorts. For cohorts including proton-treated patients, models with the highest mean AUC are reported for both constant- and variable-RBE-weighted doses (constant RBE: 1.0 for photons and 1.1 for protons; variable RBE: 1.0 for photons and RBE = 1 + m·$LET_d$ for protons, where m = 0.10 µm/keV was derived from voxel-level analyses). Regression coefficients ($\beta_i$) are given in Gy(RBE)$^{-1}$ for $D_x$ variables and cm$^{-3}$ for $V_x$ variables.

**Photon-proton**

| RBE | OAR | Parameter | AUC | (95%-CI) | $-\beta_0$ | (95%-CI) | $\beta_1$ | (95%-CI) |
|---|---|---|---|---|---|---|---|---|
| constant | Brain | $V_{50}$ | 0.65 | (0.52-0.78) | 1.43 | (2.19-0.68) | 0.0054 | (0.0022-0.0086) |
| constant | Brain-CTV | $V_{45}$ | 0.59 | (0.45-0.73) | 0.97 | (1.71-0.24) | 0.0047 | (0.000014-0.0095) |
| constant | PVR | $D_2$ | 0.64 | (0.52-0.77) | 6.12 | (10.00-2.05) | 0.10 | (0.032-0.17) |
| variable | Brain | $D_{15}$ | 0.76 | (0.64-0.87) | 13.00 | (18.00-8.12) | 0.20 | (0.12-0.27) |
| variable | Brain-CTV | $D_{11}$ | 0.71 | (0.58-0.83) | 9.60 | (14.00-5.38) | 0.14 | (0.079-0.21) |
| variable | PVR | $D_2$ | 0.78 | (0.67-0.89) | 8.72 | (12.00-5.01) | 0.13 | (0.075-0.19) |

**Photon**

| RBE | OAR | Parameter | AUC | (95%-CI) | $-\beta_0$ | (95%-CI) | $\beta_1$ | (95%-CI) |
|---|---|---|---|---|---|---|---|---|
| constant | Brain | $V_{10}$ | 0.75 | (0.56-0.93) | 5.83 | (9.49-2.18) | 0.0059 | (0.0018-0.0100) |
| constant | Brain-CTV | $V_{10}$ | 0.69 | (0.49-0.89) | 4.47 | (7.72-1.21) | 0.0052 | (0.00083-0.0095) |
| constant | PVR | $V_{45}$ | 0.60 | (0.34-0.86) | 1.53 | (2.85-0.21) | 0.041 | (-0.020-0.10) |

**Proton**

| RBE | OAR | Parameter | AUC | (95%-CI) | $-\beta_0$ | (95%-CI) | $\beta_1$ | (95%-CI) |
|---|---|---|---|---|---|---|---|---|
| constant | Brain | $V_{55}$ | 0.81 | (0.67-0.96) | 1.15 | (1.85-0.44) | 0.012 | (0.0059-0.018) |
| constant | Brain-CTV | $V_{55}$ | 0.78 | (0.62-0.93) | 0.82 | (1.42-0.21) | 0.024 | (0.011-0.036) |
| constant | PVR | $V_{55}$ | 0.79 | (0.63-0.95) | 0.96 | (1.63-0.30) | 0.18 | (0.084-0.29) |
| variable | Brain | $D_2$ | 0.83 | (0.69-0.97) | 24.00 | (34.00-13.00) | 0.33 | (0.19-0.48) |
| variable | Brain-CTV | $D_1$ | 0.80 | (0.66-0.95) | 17.00 | (25.00-9.21) | 0.24 | (0.13-0.35) |
| variable | PVR | $D_1$ | 0.82 | (0.68-0.97) | 15.00 | (23.00-7.87) | 0.23 | (0.12-0.33) |

*Abbreviations:* AUC, mean area under the receiver operating characteristic curve; CI, confidence interval; CTV, clinical target volume; $D_x$, minimum dose to the *x* ml of the OAR receiving the highest dose; OAR, organ at risk; PVR, periventricular region; RBE, relative biological effectiveness; $V_x$, volume of the OAR receiving more than *x* Gy(RBE).

## Derivation of the relative biological effectiveness model

The normal tissue complication probability (NTCP)–based relative biological effectiveness (RBE) model was derived from the multivariable voxel-level logistic regression described in the main manuscript.

NTCP was modeled using a logistic function,

$$NTCP = \frac{1}{1+e^{-S}}, \quad \text{(S1)}$$

with the linear predictor

$$S = \beta_0 + \beta_1 \cdot D + \beta_2 \, D \cdot LET_d \cdot proton\ treatment + \beta_3 \cdot PVR. \quad \text{(S2)}$$

Here, $\beta_i$ denote regression coefficients, D is the physical absorbed dose, $LET_d$ is the dose-averaged linear energy transfer, and PVR the binary risk factor periventricular region, defined as 4 mm margin surrounding the cerebral ventricles.

To derive the RBE expression, equal NTCP values were assumed for photon and proton irradiation at the voxel level,

$$NTCP_{photon} = NTCP_{proton}. \quad \text{(S3)}$$

Substitution of the corresponding linear predictors yields

$$\beta_0 + \beta_1 \cdot D_{proton} + \beta_2 \cdot D_{proton} \cdot LET_{d,\ proton} + \beta_3 \cdot PVR = \beta_0 + \beta_1 \cdot D_{photon} + \beta_3 \cdot PVR. \quad \text{(S4)}$$

After cancellation of identical terms, the ratio of isoeffective doses defines the NTCP-based RBE as

$$RBE_{NTCP} = \frac{D_{photon}}{D_{proton}} = 1 + \frac{\beta_2}{\beta_1} \cdot LET_{d,\ proton}. \quad \text{(S5)}$$